\documentclass[preprint,showpacs,preprintnumbers,amsmath,amssymb]{revtex4}
\usepackage{graphicx}
\usepackage{dcolumn}
\usepackage{bm}
\begin{document}
\title {Observing the spin Coulomb drag in spin-valve devices}
\author {G. Vignale\footnote{email: vignaleg@missouri.edu}}
\affiliation
{Department of Physics and Astronomy,Ê University of Missouri, Columbia,
Missouri 65211, USA}
\date{\today}
\begin{abstract}
The Coulomb interaction between electrons of opposite spin orientations in a metal or in a doped semiconductor results in a negative off-diagonal component of  the electrical resistivity matrix -- the so-called ``spin-drag resistivity".   It is generally quite difficult to separate the spin-drag contribution from more conventional mechanisms of resistivity.  In this paper I discuss two methods to accomplish this separation in a spin-valve device.
\end{abstract} 
\maketitle
It is theoretically well established~\cite{Damico00}-\cite{Flensberg01} that the Coulomb interaction between electrons in a metal or in a doped semiconductor has a deeper effect on spin-polarized currents than on ordinary spin-unpolarized ones.  The main reason for this is that the {\it difference} between the momenta of the up-spin and down-spin electrons is not conserved in a Coulomb scattering event:  the transfer of momentum between electrons of opposite spin orientations therefore provides an intrinsic  mechanism for the decay of a spin current, even in the absence of electron-impurity scattering.  This effect has been called ``spin Coulomb drag" \cite{Damico00}, or just spin drag for brevity.  Mathematically, the spin-drag effect is best described in terms of the so-called spin transresistivity $\rho_{\uparrow\downarrow}$, which is defined as follows:  Let $j_\uparrow$ and $j_\downarrow$ be the electrical currents associated with up- and down-spin electrons (we consider here for simplicity only currents in the $x$-direction and neglect spin-orbit effects), and let $E_\uparrow$, $E_\downarrow$ be the electro-chemical fields acting on the up- and down-spins respectively (The electro-chemical field $E_\sigma$ is defined as  the gradient of the electro-chemical potential $\mu_\sigma$ divided by $e$.  The electro-chemical potential itself is the sum of the true electric potential, which determines the position of the bottom of the conduction band, and the chemical potential, which determines the level of occupation of the band.)  Then, for small departures from equilibrium one has
\begin{equation}
E_{\sigma} = \sum_{\sigma'}\rho_{\sigma \sigma'}j_{\sigma'}~,
\end{equation}
where the resistivity matrix $\rho_{\sigma\sigma'}$ has the form~\cite{Damico02}
\begin{equation} \label{resistivity} \rho ={m^* \over n
e^2 \tau} 
\left(\begin{array}{cc}2+\gamma \tau
 &  -\gamma \tau \\
-\gamma \tau &
2+\gamma \tau
 \end{array} \right ).  
\end{equation}
In the above equation $\gamma$ is the spin-drag coefficient, i.e., the intrinsic relaxation rate of the spin momentum $p_\uparrow-p_\downarrow$, $\frac{1}{\tau}$ is the ordinary momentum relaxation rate due to electron impurity interactions, $m^*$ and $e$ are the band mass and the absolute value of the electron charge,  and $n$ is the total electronic density.  Eq.~(\ref{resistivity}) is valid under the assumption that the spin-flip scattering rate is negligible in comparison to $\gamma$ -- a condition that should be well satisfied except at very low temperatures~\cite{Damico03}.  We have also assumed, for simplicity, that the system is paramagnetic, i.e.,  $n_\uparrow=n_\downarrow=\frac{n}{2}$, so that $\rho_{\uparrow\uparrow}=\rho_{\downarrow\downarrow}$.  Looking at Eq.~(\ref{resistivity}) we notice an important fact:  $\rho_{\uparrow\downarrow}$ is {\it negative}, because it takes a negative electro-chemical field to prevent an up-spin current from flowing when a down-spin current is present.  On the other hand, the positivity of dissipation requires both eigenvalues of $\rho$ to be positive -- a condition that is obviously satisfied by Eq.~(\ref{resistivity}) provided  $\gamma$ is positive.  

It is clear that an experimental determination of $\gamma$ would be of great interest since the value of this quantity is controlled by many-body correlations, which are  intrinsic to the equilibrium state of the electron liquid.  The main difficulty  is that the spin transresistivity cancels out in the ordinary resistivity $\rho=\frac{\rho_{\uparrow\uparrow}+\rho_{\uparrow \downarrow}}{2}$, so one has to devise an experiment that is somehow sensitive to the ``spin resistivity" $\rho_{spin} = \frac{\rho_{\uparrow\uparrow}-\rho_{\uparrow \downarrow}}{2}=\rho (1+\gamma\tau)$.   An obvious way to proceed,  first proposed in Ref.~\cite{Damico00}, is  to measure the electro-chemical potential drop in the down-spin component when an up-spin-polarized current  is driven into the semiconductor via highly spin-polarized ferromagnetic electrodes (spin injectors).  
\begin{figure}
\label{Figure1}
\includegraphics[width=7cm]{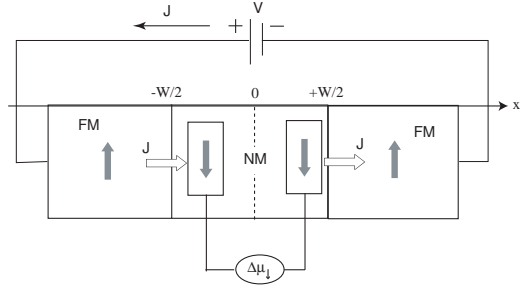}
\caption {Spin-valve device for the measurement of the spin drag effect. A predominantly up-spin current is injected in a non-magnetic (NM) semiconductor via ferromagnetic (FM) electrodes between which a potential difference $V$ is applied.  The voltage probes are polarized oppositely to the injectors and therefore measure the down-spin electro-chemical potential.}
\vspace{0pt}
\end{figure}
In the limit that  the level of spin-polarization $p$ of the ferromagnetic electrodes is $100\%$ ``up", and the spin diffusion length of the semiconductor  is much larger than its geometrical length,  the injected current is entirely in the up-spin component.   Under these conditions,  the electro-chemical potential ``drop" for down-spins will be negative, if $\gamma$ is a finite positive quantity, and would vanish if $\gamma =0$.
If, on the other hand, the polarization of the electrodes is less than $100\%$ then the down-spin electro-chemical potential drop may remain positive, foiling our attempts to detect and measure $\gamma$.
 
Thus, a very important question concerning this proposal is: how large should the polarization of the electrodes be so that one may observe a {\it negative} down-spin electro-chemical potential difference (as opposed to the trivially positive electric potential difference)?

This paper is largely devoted to providing a sharp answer to this question.  It will be shown  (see Eqs.~(\ref{Eupdownfm}) and (\ref{pcritical}) below)  that as $p$  increases from $0$ to $100\%$ there is a critical value of $p$, given by  $p_c=\frac{1}{1+\gamma\tau}$,  at which the electro-chemical potential drop for down-spins switches from positive to negative.  Measuring $p_c$ amounts therefore to a measurement of  $\gamma \tau$.  The experiment could be carried out in a three-layer spin-valve structure~\cite{Baibich88,Valet93,Schmidt00,Schmidt01,Awschalom02}, such as the one shown in Fig.~1.  The two  electrodes/spin injectors  could be made out of a large-$g$-factor II-VI semiconductor, e.g.  Be$_x$Mn$_y$Zn$_{1-x-y}$Se, where $g \sim 100$~\cite{Fielderling99}, which can be completely polarized by the application of a modest magnetic field.    These electrodes are used to inject a spin-polarized current into a  nonmagnetic (NM) lightly doped semiconductor (e.g., GaAs) and the total resistance across the electrodes is measured.  The main physical assumptions underlying the proposed measurement are as follows:
\begin{enumerate}
\item The spin drag effect is important only in the nonmagnetic semiconductor (GaAs), where the density of carriers is low.  This is because it is theoretically  well established that the spin drag increases in magnitude as the density of the electrons decreases~\cite{Damico00}.  
\item The magnetic field, which is needed to spin-polarize the electrodes,  has a negligible effect on the electronic states in the non-magnetic semiconductor, in which the $g$-factor is small.  
\item   The spin-resolved conductivities  of the electrodes $\sigma^f_{\uparrow\uparrow}$ and $\sigma^f_{\downarrow\downarrow}$  scale in proportion to the corresponding electron densities, i.e.,  $\sigma^f_{\uparrow\uparrow}= \frac{1+p}{2}\sigma^f$ and $\sigma^f_{\downarrow\downarrow}= \frac{1-p}{2}\sigma^f$, where $\sigma^f$ is the total conductivity of the homogeneous ferromagnet.  (Of course $\sigma^f$ itself may slightly depend on $p$:  this question will be discussed below.)
 \end{enumerate}

 The analysis is based on the equation for the electro-chemical potentials derived in Ref.~\cite{Damico03.2}.  In the one-dimensional geometry of Fig.1 this takes the form 
\begin {equation} \label{muequation}
\frac{d^2 \mu_\sigma(x)}{d x^2}= \sum_{\sigma'}M_{\sigma \sigma'}\mu_{\sigma'}~,
\end{equation}
where the $2 \times 2$ matrix $M_{\sigma \sigma'}$ is, for our purposes, completely specified by its  right eigenvectors, namely
\begin{equation}
\left(\begin{array}{l}1\\ 1 \end{array}\right)
\end{equation}
 (the charge mode)   with eigenvalue $0$, and
\begin{equation}
\left(\begin{array}{l}~~1\\ -1\end{array}\right)
\end{equation}
 (the spin mode)   with eigenvalue $\frac{1}{L^2}$, where $L$ is the spin diffusion length.  The solution of Eq.~(\ref{muequation}) is straightforward.   To make the best use  of symmetry we assume that the semiconductor layer extends from $x=-W/2$ to $x=W/2$.  The electro-chemical potentials are then odd functions of $x$ [$\mu_\alpha(x)=-\mu_\alpha(-x)$], and we can focus only on the region $x<0$.   In this region we write
 \begin{equation}\label{muupdownfm}
\left(\begin{array}{c} \mu_\uparrow  \\ \mu_\downarrow\end{array}\right) = \left \{\begin{array}{c} \frac{e J W}{\sigma^f}\left[\left[-\frac{C_0}{2} + \left (\frac{1}{2} +\frac{x}{W} \right)   
\right] \left( \begin{array}{c} 1\\ 1  \end{array}\right) 
+ 2 C_1
e^{\frac{W/2+x}{L^f}} \left( \begin{array}{c} ~(1+p)^{-1} \\ -(1-p)^{-1}\end{array}\right) \right]~,~~~~ x<-\frac{W}{2} ~\\
\frac{e J W}{\sigma^s} \left[\frac{x}{W} \left( \begin{array}{c} 1\\ 1  \end{array}\right)+ 2C_2
\sinh{\left(\frac{x}{L^s}\right)} \left( \begin{array}{c} ~1\\ -1 \end{array}\right)\right]~,~~~~~~~~~~~~~~~~~~~~~~~-\frac{W}{2}\leq x\leq 0~\end{array}\right.~,
\end{equation}
where  $J$ is the charge current, $\sigma^f$ and $\sigma^s$  are the conductivities of the electrodes and of the semiconductor, and $L^f$ and $L^s$ the spin diffusion lengths in the electrodes and in the semiconductor, respectively.     Notice that the continuity of the charge current, $J$, is  already built in Eq.~(\ref{muupdownfm}).  The three constants $C_0$, $C_1$, and $C_2$ are determined from the continuity of the two electro-chemical potentials and of the spin current $j_\uparrow(x) - j_\downarrow(x)$ at $x = -W/2$.   Their explicit forms are easily found to be
\begin{eqnarray}\label{constants}
C_0 &=&\frac{\sigma^f}{\sigma^s}+ \frac{2 p^2}{{\cal D}} \sinh{\left( \frac{W}{2L^s}\right)}~, \nonumber \\
C_1&=&- \frac{p(1-p^2)}{2{\cal D}}\sinh{\left( \frac{W}{2L^s}\right)}~, \nonumber \\
C_2&=&\frac{p\sigma^s}{2\sigma^f {\cal D}}~,
\end{eqnarray}
with
\begin{equation}\label{denominator}
{\cal D}=\frac{W \left(1-p^2\right)}{L^f}\sinh\left( \frac{W}{2L^s}\right)+ \frac{W\sigma^s}{L^s\sigma^f} \frac{1}{1+\gamma\tau}
\cosh\left( \frac{W}{2L^s}\right)~.
\end{equation}
As mentioned above, the solution for $x>0$ is obtained by means of the symmetry relation $\mu_\sigma(x)=-\mu_\sigma(-x)$.

The behavior of the solution (expressed in units of $eJW/\sigma^s$) is shown in Fig. 2.   
\begin{figure}
\label{Figure2}
\includegraphics[width=10cm]{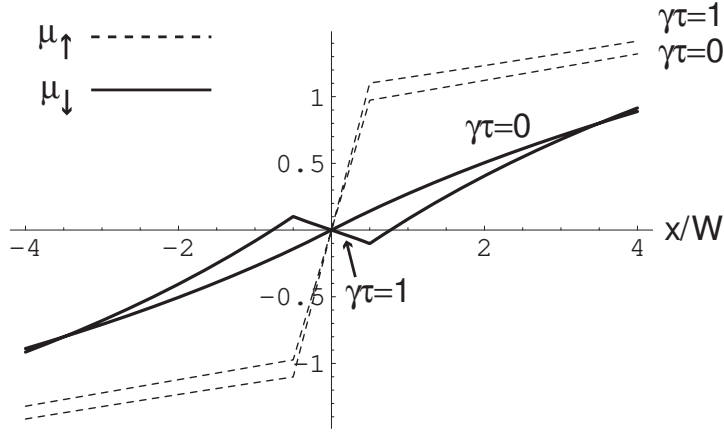}
\caption {The behavior of the electrochemical potentials $\mu_{\uparrow}$ (dashed lines)  and $\mu_{\downarrow}$ (solid lines) calculated from Eq.~(\ref{muupdownfm}) for $\gamma \tau = 0$ and $\gamma \tau =1$ and expressed in units of $\frac{e J W}{\sigma^s}$ in the parallel-electrodes configuration.  The semiconductor is in the region $-\frac{1}{2} \leq \frac{x}{W} \leq \frac{1}{2}$.  The other parameters are $p=90\%$, $\frac{\sigma^f}{\sigma^s}=10$, $\frac{L^s}{W}=2$, and $\frac{L^f}{W}=4$.  Notice the negative slope of $\mu_{\downarrow}(x)$ at $x=0$ when $\gamma \tau=1$: such a negative slope is an unmistakable signature of the spin drag.}
\vspace{0pt}
\end{figure}
Basically, we observe an accumulation of down-spin electrons and a corresponding depletion of up-spin electrons (i.e., $\mu_\downarrow>\mu_\uparrow$) at the left interface, where up-spin electrons are injected.  The opposite happens at the left interface, where up-spin electrons are extracted.  These spin accumulations  effectively create a diffusion barrier, which increases the resistance and reduces the efficiency of spin-current injection.
Under the assumption $L^s\gg W$ the electro-chemical fields, defined as the slopes of the elctrochemical potentials  divided by $e$,  are nearly exactly uniform in the nonmagnetic region and their values are given by
\begin{eqnarray}\label{Eupdownfm}
E_{\uparrow}(0)&=&\frac{J}{\sigma^s}+\frac{JWp}{L^s\sigma^f{\cal D}}\nonumber\\
E_{\downarrow}(0)&=&\frac{J}{\sigma^s}-\frac{JWp}{L^s\sigma^f{\cal D}}~.\end{eqnarray}

Notice that $E_\downarrow$ is always smaller than  $E_\uparrow$ and would tend to zero for $p\to 100\%$ in the absence of the spin drag effect.  This is because as the polarization of the electrodes increases, the down-spin component of the current must decrease:  in the absence of spin drag this would imply that a gradient in chemical potential of down-spin electrons must be present to balance the electric field, resulting in $E_\downarrow \approx 0$.  The spin drag upsets this balance.  It is now necessary to have a finite, {\it negative} $E_\downarrow$ in order to balance the momentum transfer from up- to down-spin electrons.  The change in sign in $E_\downarrow$ is an unmistakable signature of the spin Coulomb drag and occurs when the spin polarization of  the electrodes exceeds the critical value
\begin{eqnarray}\label{pcritical}
p_c&=&\frac{\sqrt{1+4\alpha^2\sinh^2\left(\frac{W}{2L^s}\right)+2\alpha\sinh \left(\frac{W}{L^s}\right)\frac{1}{1+\gamma\tau}}-1}{2 \alpha \sinh\left(\frac{W}{2L^s}\right)}
\nonumber\\
&\stackrel{W\ll L^s}{\simeq}&\frac{1}{1+\gamma\tau}~,
 \end{eqnarray}
 where $\alpha\equiv \frac{L^s\sigma^f}{L^f \sigma^s}$ (notice that $p_c>1$ for $\gamma=0$, that is, in the absence of spin drag.  For $\gamma \tau=1$ with the parameters of Fig. 2 we have $p_c \approx 0.85$).  Thus by measuring the value of $p$ at which $E_\downarrow$ changes sign one can determine $\gamma \tau$.

The main drawback of such an experimental design (which is conceptually analogous to the design of the Coulomb drag measurement in bilayer systems~\cite{Gramila91}) is the need to establish separate  electrical contacts for the up- and down-spin electrons.  This could be accomplished by the introduction of ferromagnetic voltage probes, polarized oppositely to the current leads.   Unfortunately,  such ``probes" are technically difficult to implant  and complicate the analysis of the experiment,  for they disturb the equilibrium distribution of the spin in the semiconductor.  For this reason I now describe what should be a  simpler method to determine the quantity $1+\gamma \tau$.     The idea of the measurement is simply to compare the total resistance $R$ of the circuit at $p=0$ (i.e., for unpolarized electrodes) and  $p=1$ (i.e., $100\%$ spin-polarized electrodes).  No spin-polarized voltage probes are required. We assume that the homogeneous  resistance of the electrodes and the external wires (denoted by $R_c$ for brevity)  is  small compared to the resistance of the non-magnetic semiconductor.  The polarization dependence of $R_c$ presumably amounts to an even smaller correction.  
At $p=0$ the total resistance is thus essentially equal to the ordinary resistance of the semiconductor: the spin-drag effect is invisible here.  At $p=1$,  on the other hand, the resistance depends very much on whether there is spin drag or not.  If the spin drag were absent,  then the resistance would be {\it twice} the ordinary resistance of the semiconductor, because only one of the two spin channels is open to conduction.  In the presence of spin drag the flow of the up-spin current is hindered by collisions with down-spin electrons, which are stationary on the average: as a result, the resistance of the conductor  becomes more  than twice the ordinary resistance -- in fact we will show that it is $2+\gamma \tau$ times the ordinary resistance.  Thus, by taking the difference $R(p=1)-R(p=0)$ and dividing it by $R(p=0)$ we arrive at an experimental determination of $\gamma \tau$.  It should be noted that in taking the difference $R(p=1)-R(p=0)$ the  resistance $R_c$  of the wires and the electrodes largely cancels out, except for its polarization-dependent component, which we feel justified in neglecting.   Furthermore, this determination does not depend on the value of the spin diffusion length in the semiconductor, $L^s$, provided the latter is much larger than the length  of the semiconductor itself, $W$ -- a condition that should not be too difficult to satisfy in practice.   Likewise, the value of the spin-diffusion length in the electrodes, $L^f$, is essentially  irrelevant as long as the potential drop  is measured  between points that are much farther than a distance $L^f$ from the FM-NM interfaces.


\begin{figure}
\label{Figure3}
\includegraphics[width=10cm]{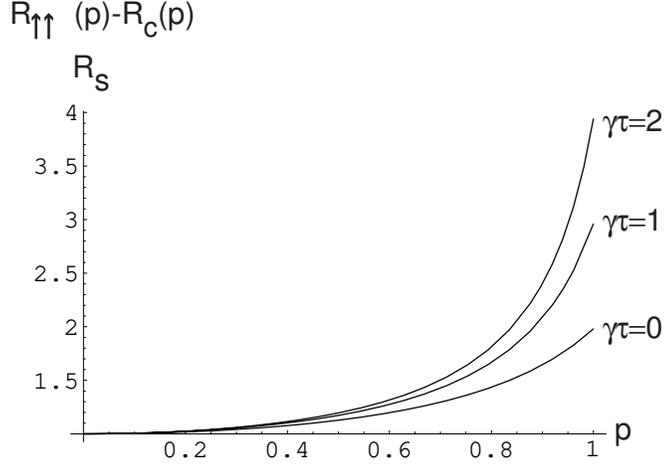}
\caption {Behavior of the parallel-electrodes magneto-resistance $R_{parallel}(p)-R_c(p)$ (in units of the ordinary resistance of the semiconductor, $R_s$) as a function of $p$ for  $\gamma \tau=0$, $1$, and $2$.  Notice the sharp enhancement caused by the spin-drag resistivity for values of $p$ close to $100\%$.  As explained in the text,  this can be used to determine $\gamma \tau$. }
\vspace{0pt}
\end{figure}

Far from the FM-NM interfaces (more precisely, at a distance much larger than $L^f$) the electro-chemical potentials of the two spin orientations tend to common values  $\mu_- = -\frac{eJW(C_0-1)}{2 \sigma^f} +\frac{e J x}{\sigma^f}$ for $x \to -\infty$ and  $\mu_+ = +\frac{eJW(C_0-1)}{2 \sigma^f} +\frac{e J x}{\sigma^f}$  for $x \to +\infty$.   The difference between these two asymptotic behaviors is $e$ times the voltage drop due to the presence of the semiconductor layer.  Hence, the resistance of our device (per unit cross-sectional area)  is given by
\begin{equation}\label{Ra}
R_{parallel}(p) =  R_c(p) + R^s+\frac{2 W p^2}{{\cal D} \sigma^f} \sinh{\left( \frac{W}{2L^s}\right)}~,
\end{equation} 
where $R_c(p)$ is the combined resistance of the electrodes and the external wires, and $R^s=\frac{W}{\sigma^s}$  is the ordinary resistance of the  semiconductor. The last term on the right hand side of this equation arises from the spin accumulations at the interfaces between the electrodes and the  semiconductor.  Fig. 3 shows the behavior of  the key quantity $R_{parallel}(p)-R_c(p)$ as a function of  $p$.  It increases from $R_s$ at $p=0$ to $R^s(2+\gamma \tau)$ at $p=1$.  Interestingly most of the change occurs in the region of $p$ close to 1.  This can be exploited  to reduce the undesired effect of the $p$-dependence of $R_c$.  Namely, rather than considering the change in  resistance from $p=0$ to $p=1$, it may be sufficient to consider the change from say $p=0.5$ to $p=1.0$ with correspondingly less variation in $R_c$.  
Notice that heoretical calculations of $\gamma$ as a function of temperature and electronic density can be found in Refs.~\cite{Damico00}-\cite{Damico03}.  The temperature dependence of $\gamma$ is particularly interesting as it exhibits a characteristic broad maximum at about the degeneracy temperature of the carriers in the semiconductor.

For completeness, let us now see what happens  in the antiparallel-electrodes configuration.  In this case, the electro-chemical potentials obey the symmetry relation
$\mu_\alpha(x)=-\mu_{\bar \alpha}(-x)$ and it is easy to see that the new solution is now obtained from the parallel case solution simply by interchanging the quantities $\sinh{\left( \frac{W}{2L^s}\right)}$ and
$\cosh{\left( \frac{W}{2L^s}\right)}$.   More precisely, the solution for $x<0$ takes the form
\begin{equation}\label{muupdownafm}
\left(\begin{array}{c} \mu_\uparrow  \\ \mu_\downarrow\end{array}\right) = \left \{\begin{array}{c} \frac{e J W}{\sigma^f}\left\{\left[-\frac{C_0'}{2} + \left (\frac{1}{2} +\frac{x}{W} \right)   
\right] \left( \begin{array}{c} 1\\ 1  \end{array}\right) 
+ 2 C_1'
e^{\frac{W/2+x}{L^f}} \left( \begin{array}{c} ~(1+p)^{-1} \\ -(1-p)^{-1}\end{array}\right) \right\}~,~~~~ x<-\frac{W}{2} ~\\
\frac{e J W}{\sigma^s} \left\{\frac{x}{W} \left( \begin{array}{c} 1\\ 1  \end{array}\right)-2C_2'
\cosh{\left(\frac{x}{L^s}\right)} \left( \begin{array}{c} ~1\\ -1 \end{array}\right)\right\}~,~~~~~~~~~~~~~~~~~~~~~~~-\frac{W}{2}\leq x\leq 0~\end{array}\right.~,
\end{equation}
where  the constants $C_0'- C_2'$ are given by
\begin{eqnarray}\label{constants}
C_0' &=&\frac{\sigma^f}{\sigma^s}+ \frac{2 p^2}{{\cal D}'} \cosh{\left( \frac{W}{2L^s}\right)}~, \nonumber \\
C_1'&=&- \frac{p(1-p^2)}{2{\cal D}'}\cosh{\left( \frac{W}{2L^s}\right)}~, \nonumber \\
C_2'&=&\frac{p\sigma^s}{2\sigma^f {\cal D}'}~,
\end{eqnarray}
and
\begin{equation}\label{denominator}
{\cal D}'=\frac{W \left(1-p^2\right)}{L^f}\cosh\left( \frac{W}{2L^s}\right)+ \frac{W\sigma^s}{L^s\sigma^f} \frac{1}{1+\gamma\tau}
\sinh\left( \frac{W}{2L^s}\right)~.
\end{equation}
The solution for $x>0$  is obtained by means of the symmetry relation $\mu_\sigma(x)=-\mu_{-\sigma}(-x)$.  A representative plot of $\mu_{\uparrow}$ and $\mu_{\downarrow}$ is shown in Fig. 4.
\begin{figure}
\includegraphics[width=10cm]{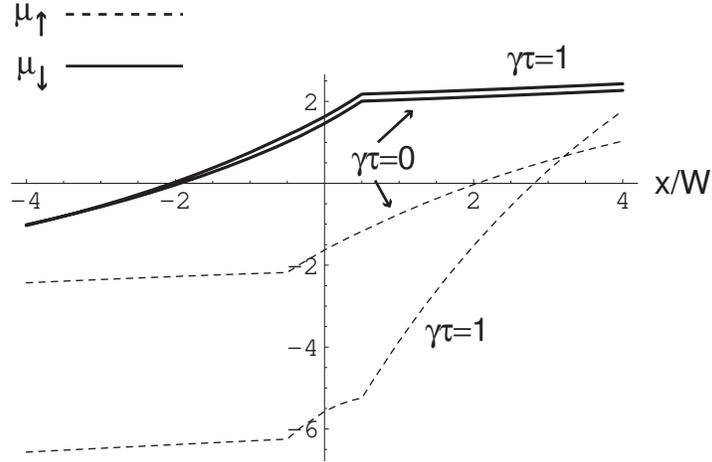}
\caption {The behavior of the electrochemical potentials $\mu_{\uparrow}$ (dashed lines)  and $\mu_{\downarrow}$ (solid lines) calculated from Eq.~(\ref{muupdownafm}) for $\gamma \tau = 0$ and $\gamma \tau =1$ and expressed in units of $\frac{e J W}{\sigma^s}$ in the antiparallel-electrodes configuration.  The parameters are the same as in the caption of Fig. 2, namely  $p=90\%$, $\frac{\sigma^f}{\sigma^s}=10$, $\frac{L^s}{W}=2$, and $\frac{L^f}{W}=4$.}
\vspace{0pt}
\label{Figure4}
\end{figure}

\begin{figure}
\label{Figure5}
\includegraphics[width=10cm]{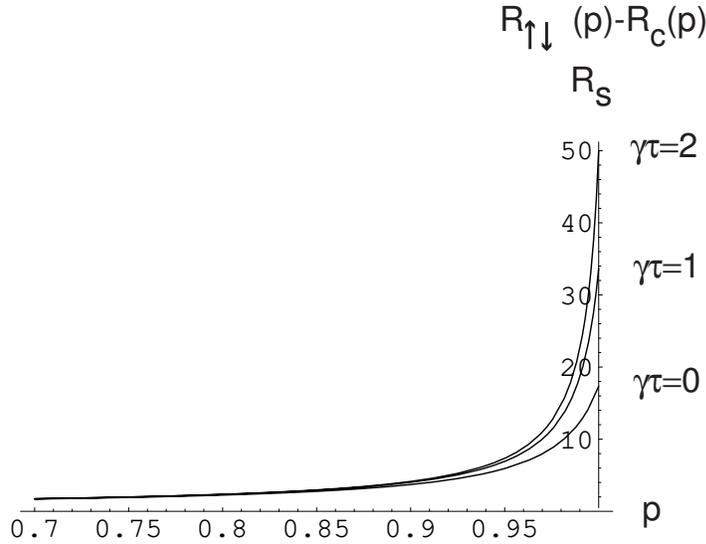}
\caption {Behavior of the antiparallel-electrodes magneto-resistance $R_{parallel}(p)-R_c(p)$ (in units of the ordinary resistance of the semiconductor, $R_s$) as a function of $p$ for  $\gamma \tau=0$, $1$, and $2$.  The enhancement caused by the spin-drag resistivity for values of $p$ close to $100\%$ is now amplified  by a factor $\left(\frac{2L^s}{W}\right)^2$ ($= 16$ in the present case).  As explained in the text,  this can be used to mesaure $\frac{L^s}{W}$ once  $\gamma \tau$ is known. }
\vspace{0pt}
\end{figure}

We  can calculate the resistance of the antiparallel-electrodes configuration in precisely the same way as in the parallel-electrodes case.  The result is
\begin{equation}\label{Rb}
R_{antiparallel} =  R_c(p) + R^s+\frac{2 W p^2}{{\cal D}' \sigma^f} \sinh{\left( \frac{W}{2L^s}\right)}~,
\end{equation} 
and the quantity $R_{antiparallel}(p) -  R_c(p)$ is plotted vs $p$ in Fig. 4.  The resistance of this configuration is of course much larger than that of the parallel configuration (this is the well known GMR effect) and it is easy to see that in the limit $p \to 1$ it tends to  to $R^s(1+\gamma \tau)\left(\frac{2L^s}{W}\right)^2$.  Notice that the Coulomb enhancement in this configuration is  very sharply confined to the region of  $p\sim1$.   The results of the above calculation can be used to determine $\frac{L^s}{W}$, once $\gamma \tau$ has been determined from the measurement of the resistance in the parallel-electrodes configuration.

In summary I have theoretically analyzed in this paper two methods to measure  the spin drag coefficient of a non-magnetic semiconductor sandwiched between highly spin-polarized ferromagnetic electrodes.  The first method  builds upon  the {\it gedanken experiment} proposed in Ref.~\cite{Damico00} showing that an unambiguous {\it qualitative} signature of the spin drag effect occurs when the spin polarization of the ferromagnetic electrodes exceed the critical value $p_c \simeq \frac{1}{1+\gamma \tau}$.    In the second, more {\it quantitative}  method one simply measures the extra resistance introduced by the relative motion of the up-spin and down-spin electrons in the semiconductor region of a basic spin-valve device.   It is hoped that these discussions will encourage further experimental work aimed at the observation of the spin Coulomb drag.

The author gratefully acknowledges the hospitality of the Center for Nanoscience of the University of Science and Technology of China,  Hefei, China, where much of this work was completed, and support  from  NSF Grant No. DMR-0313681.   Many thanks also to Irene D'Amico for many useful comments on the manuscript.

\end{document}